\documentclass{aa}  

\usepackage{natbib}

\usepackage{caption}

\usepackage[flushleft]{threeparttable}

\usepackage{graphicx}
\usepackage{txfonts}
\usepackage{float}
\usepackage{subfloat}

\usepackage{subfig}

\usepackage{gensymb}

\usepackage{url}

\makeatletter
\newcommand\footnoteref[1]{\protected@xdef\@thefnmark{\ref{#1}}\@footnotemark}
\makeatother

\begin{document}

   \title{VLBA+GBT observations of the COSMOS field and radio source counts at 1.4 GHz\thanks{Table~\ref{table:agncatgbt} and Table~\ref{table:mwgbt} will be also available in electronic form at the CDS.}}

   \author{N. Herrera Ruiz\inst{1}
          \and
          E. Middelberg\inst{1}
          \and
          A. Deller\inst{2,3}
          \and
          V. Smol{\v c}i{\'c}\inst{4}
          \and
          R. P. Norris\inst{5,6}
          \and
          M. Novak\inst{4}
          \and
          I. Delvecchio\inst{4}
          \and
          P. N. Best\inst{7}
          \and
          E. Schinnerer\inst{8}
          \and
          E. Momjian\inst{9}
          \and
          R.-J. Dettmar\inst{1}
          \and
          W. Brisken\inst{9}
          \and
          A. M. Koekemoer\inst{10}
          \and
          N. Z. Scoville\inst{11}
          }

   \institute{Astronomisches Institut, Ruhr-Universit\"at Bochum, 
              Universit\"atsstrasse 150, 44801 Bochum, Germany\\
              \email{herrera@astro.rub.de}
         \and The Netherlands Institute for Radio Astronomy (ASTRON), 7991 Dwingeloo, The Netherlands
         \and Centre for Astrophysics and Supercomputing, Swinburne University of Technology, PO Box 218, Hawthorn, VIC 3122, Australia
         \and Department of Physics, Faculty of Science, University of Zagreb, Bijeni\v{c}ka cesta 32, 10000 Zagreb, Croatia
         \and CSIRO Australia Telescope National Facility, PO Box 76, Epping, NSW 1710, Australia
         \and Western Sydney University, Locked Bag 1797, Penrith South, NSW 1797, Australia
         \and SUPA, Institute for Astronomy, Royal Observatory Edinburgh, Blackford Hill, Edinburgh EH9 3HJ, UK
         \and MPI for Astronomy, K\"onigstuhl 17, 69117 Heidelberg, Germany
         \and National Radio Astronomy Observatory, PO Box O, Socorro, NM 87801, USA
         \and Space Telescope Science Institute, 3700 San Martin Drive, Baltimore, MD 21218, USA
         \and California Institute of Technology, MC 249-17, 1200 East California Boulevard, Pasadena, CA 91125, USA
             }

   \date{Received March 6, 2018; accepted April 18, 2018} %15 September 1996 / Accepted 16 March 1997}

\abstract{We present very long baseline interferometry (VLBI) observations of 179 radio sources in the COSMOS field with extremely high sensitivity using the Green Bank Telescope (GBT) together with the Very Long Baseline Array (VLBA) (VLBA+GBT) at 1.4\,GHz, to explore the faint radio population in the flux density regime of tens of $\mu$Jy. Here, the identification of active galactic nuclei (AGN) is based on the VLBI detection of the source, i.e., it is independent of X-ray or infrared properties. The milli-arcsecond resolution provided by the VLBI technique implies that the detected sources must be compact and have large brightness temperatures, and therefore they are most likely AGN (when the host galaxy is located at z$\geq$0.1). On the other hand, this technique allows us to only positively identify when a radio-active AGN is present, i.e., we cannot affirm that there is no AGN when the source is not detected. For this reason, the number of identified AGN using VLBI should be always treated as a lower limit. We present a catalogue containing the 35 radio sources detected with the VLBA+GBT, 10 of which were not previously detected using only the VLBA. We have constructed the radio source counts at 1.4\,GHz using the samples of the VLBA and VLBA+GBT detected sources of the COSMOS field to determine a lower limit for the AGN contribution to the faint radio source population. We found an AGN contribution of >40-75\% at flux density levels between 150\,$\mu$Jy and 1\,mJy. This flux density range is characterised by the upturn of the Euclidean-normalised radio source counts, which implies a contribution of a new population. This result supports the idea that the sub-mJy radio population is composed of a significant fraction of radio-emitting AGN, rather than solely by star-forming galaxies, in agreement with previous studies.}


   \maketitle
%
%________________________________________________________________

\section{Introduction}
\label{sec:int}

The radio source count distribution \citep{ballantyne2009} has been a classic tool to determine the fraction of active galactic nuclei (AGN) of the sub-mJy radio population. Nevertheless, discrepancies exist in the analysed contributions to the faint radio population (see \citealt{padovani2016} for a thorough review of the faint radio sky). A possible reason might be the difficulty of identifying and separating AGN and star-forming galaxies. \citet{gruppioni2003} suggested that the fraction of starburst galaxies is expected to be 60\% in the $\mu$Jy regime. \citet{padovani2009} found a roughly equal contribution of star-forming galaxies and AGN to the sub-mJy sky. \citet{fomalot2006} determined that 40\% of the $\mu$Jy radio sources are dominated by AGN processes. \citet{smolcic2008} found that the faint radio population is a mixture of 30\%-40\% star-forming galaxies and 60\%-70\% AGN. \citet{simpson2006} suggested that the flattening of the radio source counts that appears below 1\,mJy when the radio source counts are normalised to an Euclidean space may be mainly due to radio-quiet AGN.

A drop in the AGN source counts at $\sim$70$\mu$Jy is indicated by simulations of the extragalactic radio continuum sky \citep{wilman2008} and a similar drop is seen in the work by \citet{padovani2011}. The flattening of the Euclidean-normalised radio source counts suggests that the processes taking place in the sub-mJy population are different from the population at larger flux densities and has mostly been associated with star-forming galaxies \citep{seymour2004}. \citet{jarvis2004} were the first to suggest that the radio-quiet AGN population, which constitutes 90\% of the AGN population, may contribute significantly to this upturn, instead of solely the low-redshift, star-forming population. In addition, various multi-wavelength studies argue that the impact of radio-quiet AGN on the sub-mJy radio sky is still significant \citep{smolcic2008, padovani2011}. \citet{padovani2015} found that star-forming galaxies become the main population of the faint radio sky only below $\sim$0.1\,mJy and \citet{smolcic2017} that the low-to-moderate radiative luminosity AGN dominate the faint radio population above $\sim$0.2\,mJy, while the fraction of star-forming galaxies increased to $\sim$60\% below $\sim$0.1\,mJy.

In \citet{herreraruiz2017} we report on a sample of 468 radio sources of the COSMOS (Cosmic Evolution Survey) field detected with the Very Long Baseline Array (VLBA) at 1.4 GHz. These sources are considered AGN since only they can reach brightness temperatures greater than 10$^{5}$\,K \citep{condon1992}, required for a detection using very long baseline interferometry (VLBI) observations (when the redshift of the host galaxy is larger than 0.1). The median redshift of the VLBA detected sources is $\sim$1. The VLBA observations provided milli-arcsecond resolution images and a 1$\sigma$ sensitivity of 10\,$\mu$Jy/beam in the central part of the field. 

The COSMOS project is aimed to probe the evolution of galaxies through cosmic time \citep{scoville2007} and it is characterized by an extraordinary multi-wavelength coverage. It has been observed with high sensitivity in more than 30 photometric optical to near-infrared bands, and it has an excellent radio coverage. It is centred at RA (J2000) = 10:00:28.6 and Dec (J2000) = +02:12:21.0 and it covers an area of 2\,deg$^{2}$.

We have observed one additional pointing of the COSMOS field with the Green Bank Telescope (GBT) together with the VLBA (hereafter VLBA+GBT) to achieve higher sensitivity to detect even fainter sources. The pointing contained 179 radio sources from the Karl G. Jansky Very Large Array (VLA) catalogue of \citet{schinnerer2010}. We have made use of the wide-field VLBI technique (\citealt{garrett1999}; \citealt{middelberg2012}) and the multi-phase centre mode \citep{deller2007,deller2011} to carry out the observations.

The GBT is the world's largest fully steerable radio telescope. It is a parabolic off-axis reflector with a 100\,meter by 110\,meter active surface, which is able to adjust the panel positions with a high accuracy to correct for the deformation of the mirror due to the gravity. It is located in the National Radio Quiet Zone (NRQZ) of the United States, where all radio transmissions are strictly restricted by law to help scientific research and minimise interferences with the radio telescopes at Green Bank (West Virginia). The enormous collecting area of the GBT, which provides superb sensitivity, its flexibility and its location within the Radio Quiet Zone make it an ideal complement to the VLBA to study the faint radio population. The expected sensitivity improvement due to the inclusion of the GBT in the VLBA observations is about a factor of 2.8\footnote{\url{http://www.evlbi.org/cgi-bin/EVNcalc}}.

We have combined our sample of VLBA-detected sources \citep{herreraruiz2017} with the additional sources detected with the VLBA+GBT, being most likely AGN, to constrain the fraction of AGN in the $\mu$Jy regime. In this paper, we present the VLBA+GBT observations (Sect.~\ref{sec:cal}) and the resulting catalogue of detected sources (Sect.~\ref{sec:cat}). We discuss the Euclidean-normalised radio source counts in Sect.~\ref{sec:res} and Sect.~\ref{sec:dis}, and we list our main conclusions in Sect.~\ref{sec:con}.

%__________________________________________________________________

\section{Observations and data calibration}
\label{sec:cal}

\subsection{Sample}

To perform self calibration, we required the presence of a strong, unresolved source close to the field centre. Within our VLBA sample \citep{herreraruiz2017}, we detected a compact core which had a VLBA flux density of 9\,mJy and was located at \mbox{RA (J2000) = 10:01:20.06} and \mbox{Dec (J2000) = +02:34:43.67}. We identified this source as C2284 in the VLBA project and we chose it to be the main in-beam calibrator in order to be able to self-calibrate the VLBA+GBT data. We targeted all known VLA radio sources from \citet{schinnerer2010} located within a radius of 9\,arcmin (the FWHM of the GBT primary beam at 1.4 GHz) from C2284. The total number of targeted sources was 179, of which 26 were previously detected with the VLBA. Table~\ref{targetcat} (Appendix~\ref{sec:appx}) contains the information of the 179 targeted sources.

\subsection{Observations}

\begin{figure}
\centering
\includegraphics[scale=0.38]{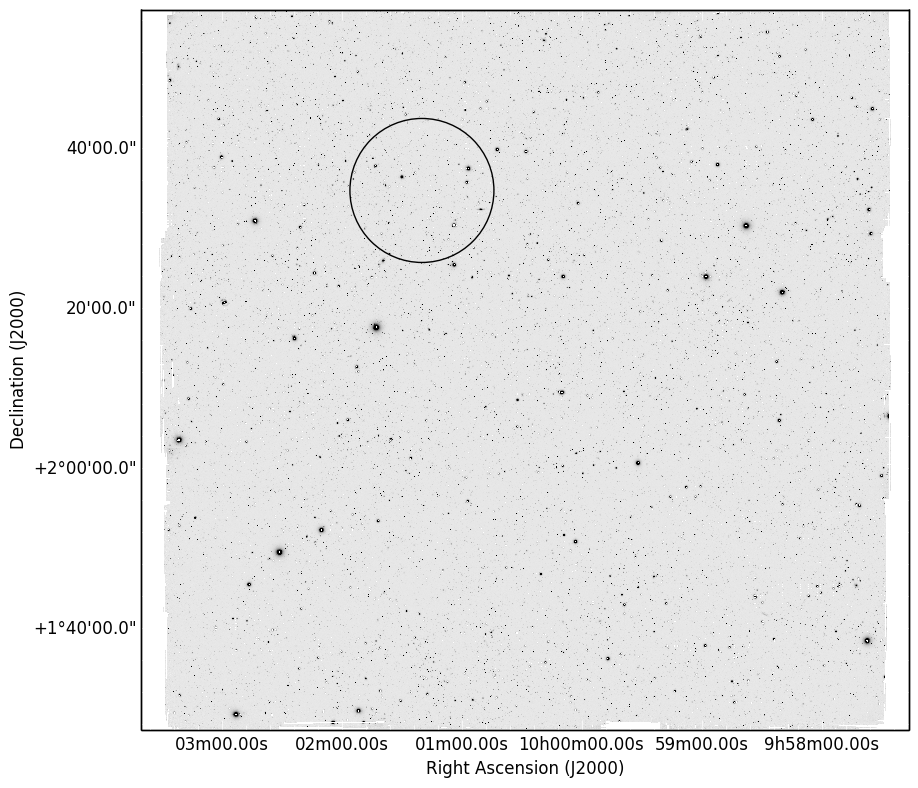}
\caption{The observed pointing using the VLBA+GBT in the COSMOS field. The black circle represents the radius within which sources were targeted (9\,arcmin). C2284 is located at the centre of the pointing. The background greyscale image is a mosaic of COSMOS Subaru 'i' band data\protect\footnotemark.}
\label{fig:designgbt}
\end{figure}

The observations of the 179 targets in the COSMOS field with the VLBA+GBT were carried out between November and December 2015. The central frequency of the observations was 1.54 GHz and we used the multi-phase centre mode of the VLBA-DiFX correlator \citep{deller2011}. The position of the observed pointing in the COSMOS field is shown in Fig.~\ref{fig:designgbt}. The pointing was observed 4 times for 6 hours, resulting in 4 individual epochs. 

Observations of the phase calibrator J1011+0106 were carried out every 26\,minutes for 1.5\,minutes, since we used the source C2284 as an in-beam calibrator. The fringe-finder 4C\,39.25 was observed every 2\,hours. Eight intermediate frequencies (IFs) with a bandwidth of 32\,MHz were observed in two circular polarizations. We requested a recording rate of 2 Gbps and a minimum number of 9 VLBA antennas together with the GBT to achieve the required sensitivity.

\footnotetext{\url{http://irsa.ipac.caltech.edu/data/COSMOS/}}

\subsection{Data calibration: Primary beam of the GBT}

We followed the same steps to calibrate the VLBA+GBT data as those followed to calibrate the VLBA data, which are explained in detail in \citet{herreraruiz2017}. In general terms, we carried out amplitude calibration using the system temperature (T$_{\mathrm{sys}}$) of the antennas and known gain curves, followed by phase calibration using the residual delay and phase measurements from the phase calibrator. The data were edited using our own flagging program. Then, we carried out the primary beam correction, and self-calibrated the data using both the in-beam calibrator and the multi-source self-calibration technique. Lastly, we combined the data collected for each source over four epochs. After calibration was complete, we created naturally-weighted images for source extraction as well as uniformly-weighted images for source flux density and position measurements. The median of the restoring beam of the naturally-weighted images was 16.1\,$\times$\,6\,mas$^{2}$ and of the uniformly-weighted images 12.3\,$\times$\,5\,mas$^{2}$, a similar resolution to that achieved with the VLBA observations (\citealt{herreraruiz2017}: 16.2\,$\times$\,7.3\,mas$^{2}$ for the naturally-weighted images and 12.4\,$\times$\,5.3\,mas$^{2}$ for the uniformly-weighted images).

The only different step was the primary beam correction because the response of the GBT primary beam was yet untested in VLBI observations. We explain the primary beam correction of the VLBA+GBT data in detail as follows.

\begin{figure}
\centering
\includegraphics[scale=0.385]{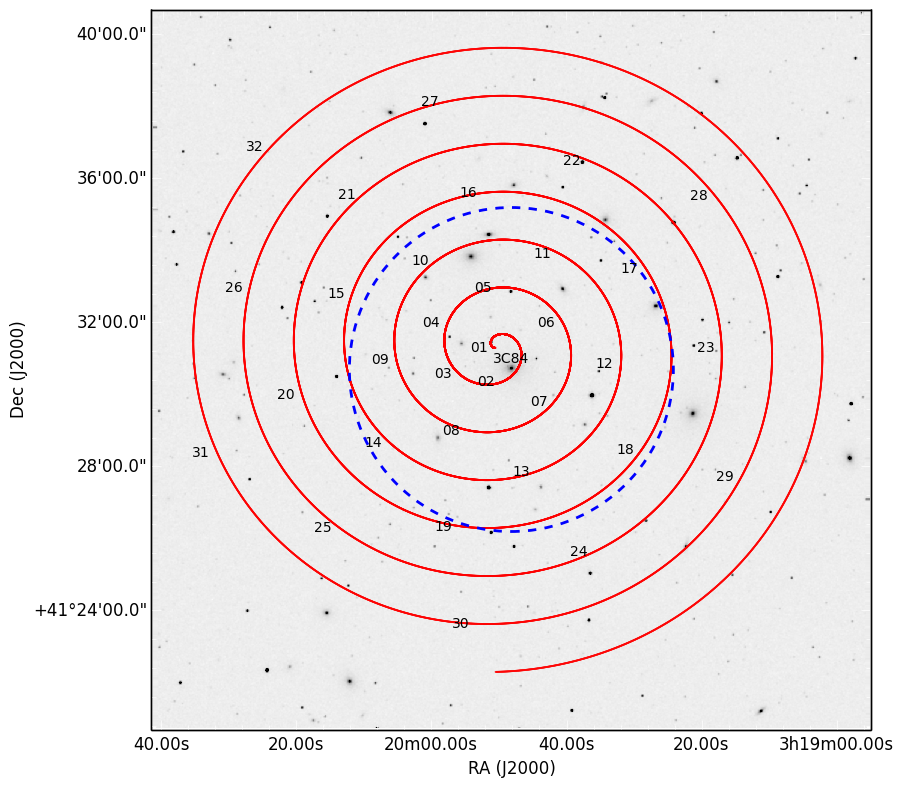}
\caption{Spiral pattern with the 32 positions observed with the GBT. The background greyscale image is a 2MASS image of 3C\,84 at $K_{s}$ band \citep{jarrett2003}. The dashed blue circle denotes the FWHM of the GBT primary beam (9\,arcmin).}
\label{fig:pospb}
\end{figure}

\citet{middelberg2013} modeled the VLBA primary beam response by observing a pattern of pointing positions around a strong radio source (3C\,84). We followed a similar method to create a model of the GBT primary beam response. For this purpose, we performed an extra 1\,hour observation. We pointed the GBT in a spiral pattern around the bright calibrator 3C\,84 (Fig.~\ref{fig:pospb}) to determine the primary beam response of the GBT. On the other hand, we pointed all the VLBA antennas to 3C\,84 continuously because the VLBA primary beam response was already known. As a result, the amplitude measured on the baselines from a VLBA antenna to the GBT would be reduced only by the primary beam response of the GBT. We placed the spiral pattern slightly offset from the central position to have the observed positions uniformly distributed within the primary beam. To measure the amplitude variation as a function of the distance from the central position (3C\,84), the correlation was carried out with 3C\,84 as the phase centre always. We observed each numbered position in the spiral pattern with the GBT for 1\,minute. We pointed the GBT back to 3C\,84 every 4\,minutes for 1.3\,minutes for phase and amplitude calibration. 

\begin{figure}
\begin{minipage}[][][t]{0.5\textwidth}
%\centering
   \includegraphics[scale=0.375]{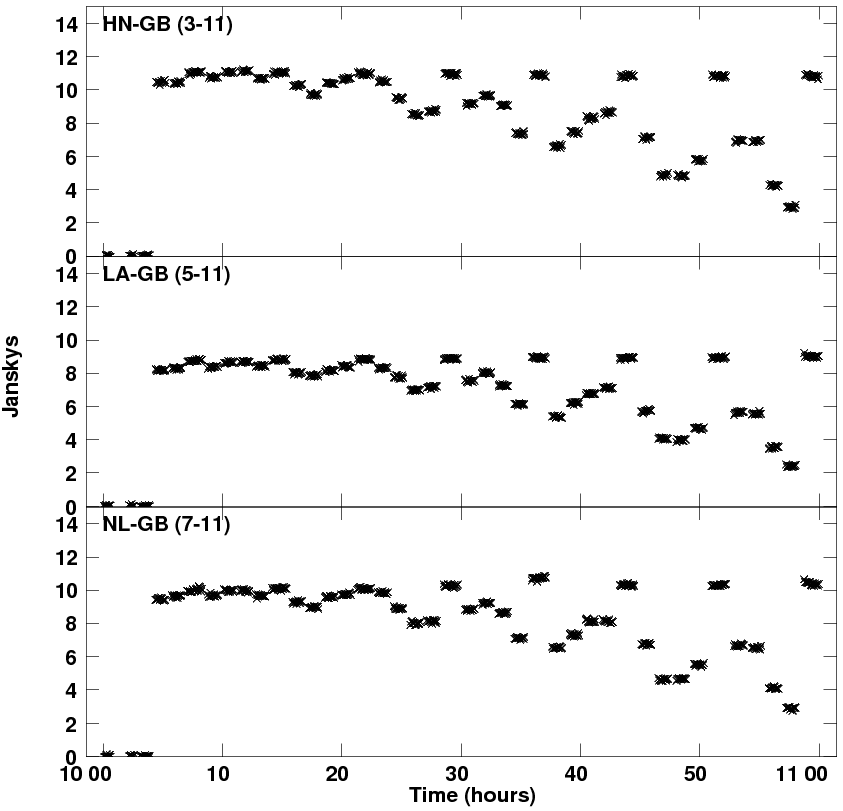}
   \includegraphics[scale=0.375]{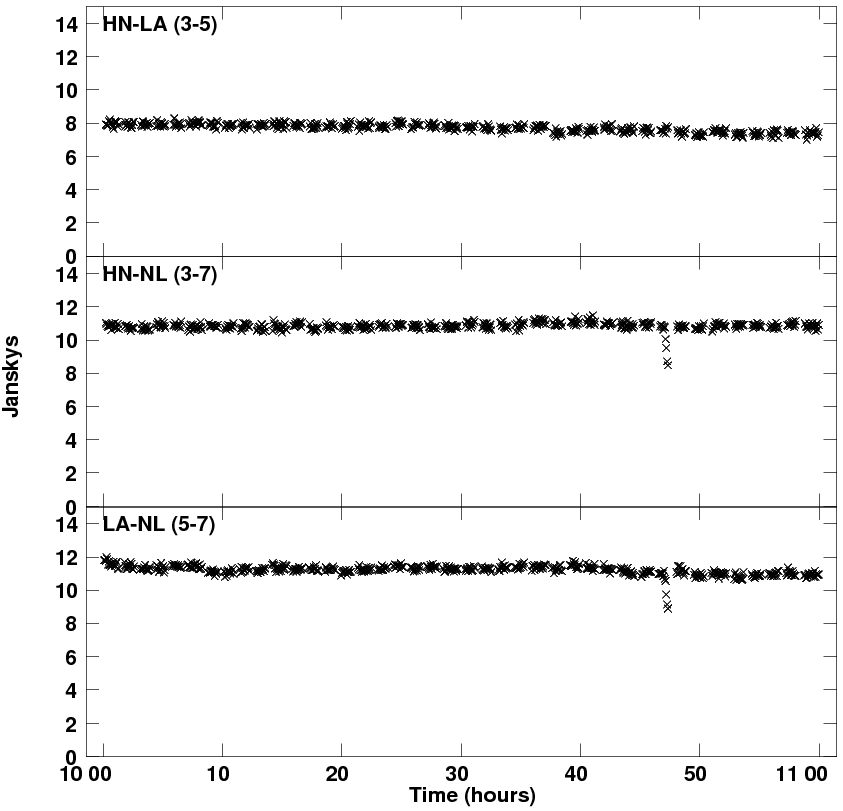}
\end{minipage}
\caption{Visibility amplitude of 3C\,84 as a function of time (first corrections and amplitude calibration applied). {\it {Top:}} Amplitude measured by the baselines between the GBT (GB) and a VLBA station (HN, LA and NL). {\it {Bottom:}} Amplitude measured only with VLBA baselines. It can be seen how the amplitude measured by the baselines between the GBT and a VLBA station changes with time because of the GBT primary beam response (see text for details). The amplitude measured only with VLBA baselines remains roughly constant.}
\label{fig:pbuncor}
\end{figure}

Figure~\ref{fig:pbuncor} represents an example of a plot showing the visibility amplitude of 3C\,84 measured by some of the baselines between the GBT (GB) and a VLBA station (HN, LA and NL) as well as the amplitude of 3C\,84 measured by baselines only between VLBA stations, as a function of time. Only first corrections and amplitude calibration were carried out at the stage when this plot was made. In the case of the amplitude measured by baselines between the GBT and a VLBA station, the variation caused by the primary beam attenuation of the GBT is noticeable. The reason for some scans being higher at later times is because some of the observed positions closer to the center of the field were observed later than others located further away due to the offset of the spiral pattern. The maximum flat scans are a result of the GBT being pointed back to 3C\,84 every 4\,minutes for calibration purposes. In the case of the baselines involving only VLBA antennas, the measured amplitude is roughly constant, as expected since the observation is relatively short.  

We performed the primary beam correction using the task CLVLB in \texttt{AIPS} (Astronomical Image Processing System, \citealt{greisen2003, fomalont1981}). This task needs the following inputs: i) the antenna number, ii) the azimuth squint, iii) the elevation squint and the frequency at which they were measured, iv) the antenna effective diameter and the frequency at which it was measured, and v) the change of effective diameter with frequency\footnote{\url{http://www.aips.nrao.edu/cgi-bin/ZXHLP2.PL?CLVLB}}. The only unknown parameter was the antenna effective diameter, $D$, of the GBT.

The power pattern of an antenna can be described by a Gaussian or by an Airy disk. To estimate $D$, we measured the variation of the amplitude as a function of the distance to 3C\,84 and we fitted an Airy disk model using the following expression:

\begin{equation}
I(\theta) = \left(\frac{2J_{1}((\pi /\lambda) D \sin \theta)}{(\pi /\lambda) D \sin \theta}\right)^{2}
\end{equation}

\noindent where $J_{1}(x)$ is the Bessel function of first order, $\theta$ is the angle between the optical axis and the observing direction, $D$ is the antenna diameter, and $\lambda$ is the observing wavelength. The Gaussian function is expressed as:

\begin{equation}
f(x) = \frac{1}{\sigma \sqrt{2 \pi}} e^{-(x-\mu)^{2}/(2\sigma^{2})}
\end{equation}

\noindent where $\sigma$ is the standard deviation and $\mu$ is the expected value. The full width at half maximum (FWHM) is given by:

\begin{equation}
\mathrm{FWHM} = 2 \sqrt{2 \ln 2}\,\sigma \approx 2.3548 \sigma
\end{equation}

\begin{figure}
\begin{minipage}[][][t]{0.5\textwidth}
%  \vspace*{\fill}
  \includegraphics[scale=0.47]{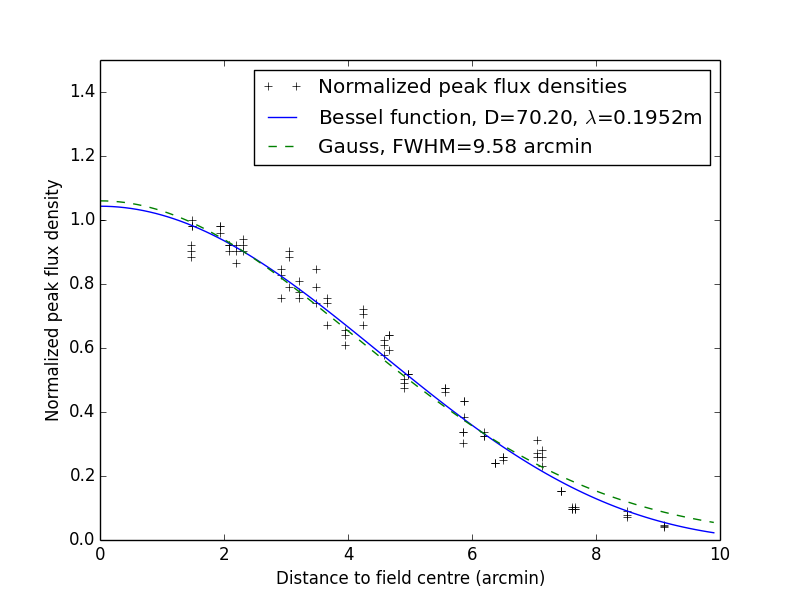}
  \includegraphics[scale=0.47]{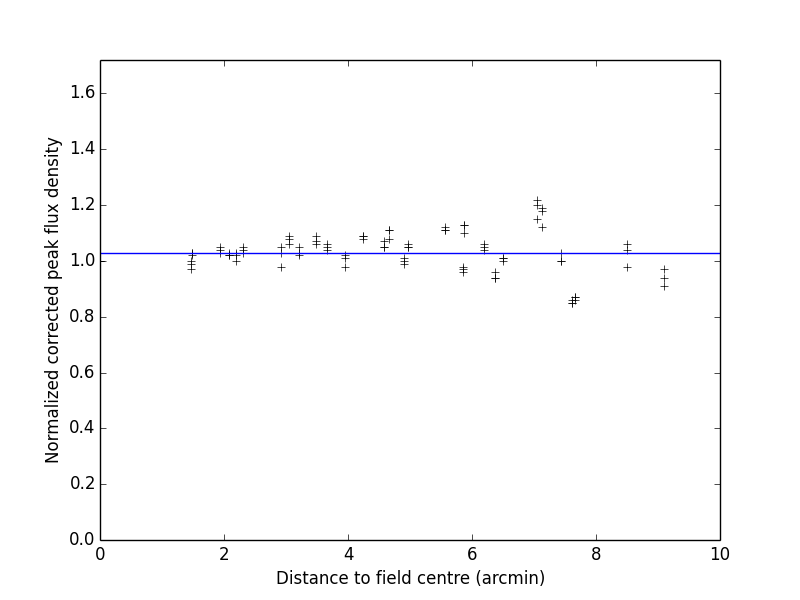}
\end{minipage}
\caption{{\it {Top:}} Normalised peak flux densities of 3C\,84 measured by three baselines between the GBT and a VLBA station as a function of the distance to the field centre. The blue solid line represents our fit to the data for an Airy disk and the green dashed line for a Gaussian. {\it {Bottom:}} The normalised peak flux densities of 3C\,84, after applying the primary beam correction, as a function of the distance to the field centre. The blue solid line represents the mean of the values, which is almost 1. The standard deviation is 0.07.}
\label{fig:pbcor}
\end{figure}

The normalised peak flux densities measured with three baselines between the GBT and a VLBA station as a function of the distance to the field centre before and after applying the primary beam correction are shown in Figure~\ref{fig:pbcor}. The Airy disk model gave an effective diameter of 70.2 meters for the GBT. The Gaussian model gave a FWHM of 9.58\,arcmin. One can see that the corrected peak flux densities after applying the primary beam correction are roughly flat, with a mean value of 1.03 and a standard deviation of 0.07. The reason for the reduced effective diameter of the GBT compared to its geometric diameter can be related to several potential loss factors like feed illumination efficiency. The slight under-illumination is intentional at the GBT in order to minimize spill-over from beyond the dish surface, which would otherwise increase T$_{\mathrm{sys}}$ and the likelihood of contamination by radio frequency interference.

\subsubsection{Sensitivity map}
\label{sec:sensmapgbt}

\begin{figure}
\centering
\includegraphics[scale=0.38]{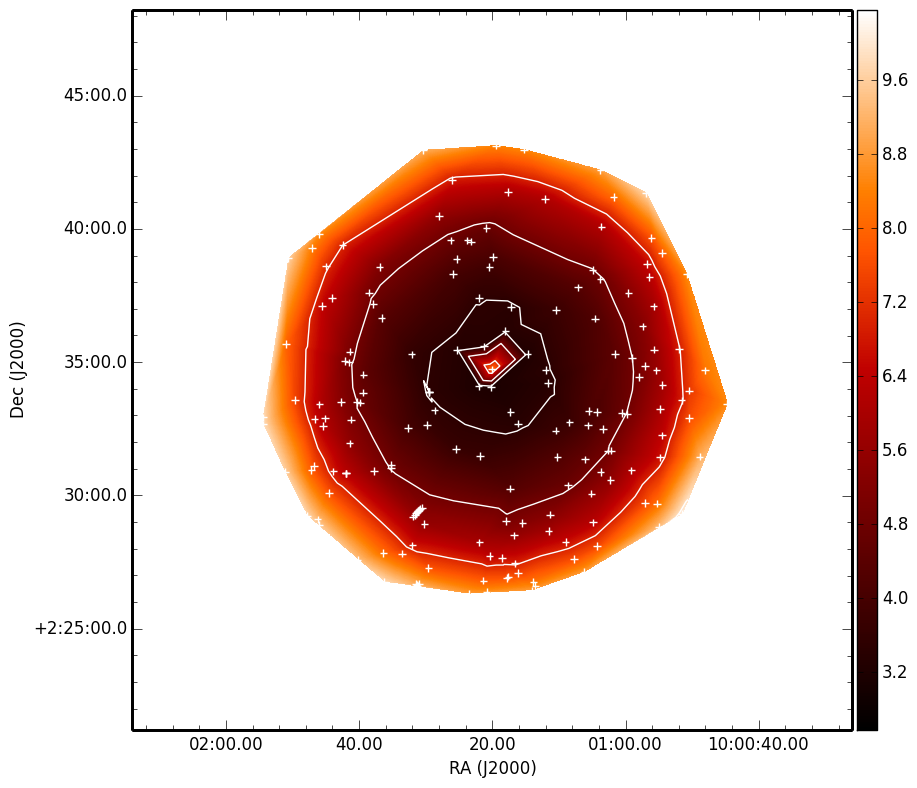}
\caption{Sensitivity map of the VLBA+GBT observations. The colour bar represents the rms noise values in $\mu$Jy/beam. The white crosses represent the target positions. White contours were drawn at 3.5, 4.9, 7 and 9.9 $\mu$Jy/beam. The high rms in the centre of the map is a result of the calibration process not being able to completely minimise the sidelobe level associated with the strong source at this location.}
\label{fig:sensitgbt}
\end{figure}

Figure~\ref{fig:sensitgbt} shows the sensitivity map of the VLBA+GBT observations. We followed the same procedure as described in \citet{herreraruiz2017} to obtain it. The rms noise level in the centre of the pointing is high because the calibration process was not able to remove entirely the sidelobe level associated with the strong radio source on which the pointing was centred.

The 100-meter diameter collecting area of the GBT and its extraordinary surface accuracy yielded an exceptional sensitivity. The achieved 1$\sigma$ rms noise level by the VLBA+GBT observations was 3\,$\mu$Jy, three times better than the sensitivity achieved by the VLBA observations \citep{herreraruiz2017}, according to our expectations.

%______________________________________________________________

\section{Catalogue}
\label{sec:cat}

\begin{table*}
\caption{Catalogue of the COSMOS VLBA+GBT detected sources (1.4 GHz).}
\label{table:agncatgbt}
\begin{threeparttable}
\centering
\begin{tabular}{lllllllllll} 
\hline\hline             
ID & COSMOSVLADP & S$_{i,VLA}$ & M & RA & Dec &  rms & S$_{p,V+G}$ & $\Delta$S$_{p,V+G}$ & S$_{i,V+G}$ & $\Delta$S$_{i,V+G}$  \\
  &   &   [$\mu$Jy]  &   & [deg] & [deg] & [$\mu$Jy/ &  [$\mu$Jy/ & [$\mu$Jy/ & [$\mu$Jy] & [$\mu$Jy]  \\
  &   &     &   &  &  & beam] &  beam] & beam] &  &  \\  
  (1) & (2) & (3) & (4) & (5) & (6) & (7) & (8) & (9) & (10) & (11)  \\
\hline
  C1886  &    J100048.89+023127.5   &   234     &  0  & 150.20371  &  2.524278   &  12.5  &  100 & 16  & 100  &   16     \\   
  C1903    &   J100050.45+023356.1   &   610     &   0   &  150.210224   &    2.565571   &    10.3          &  253        &     27     &      277     &     30       \\
  C1975   &   J100057.06+022942.9   &   123      &   0  &   150.237772    &   2.495213   &    10.3          &  103      &       15     &       103     &     15       \\ 
  C1977   &    J100057.11+023451.7   &   347     &   0   &  150.237968    &   2.581037    &   7.6          &  185       &      20   &       186   &       20       \\   
  C2010   &   J100059.05+023508.9   &   237      &   0   &  150.246091    &   2.585823     &  7.2          &  162       &      18  &         162      &    17      \\
  C2049  &    J100101.81+024111.7   &   169    &   0   &  150.257517   &    2.686597   &    10.4          &  75     &        13     &      110     &     15        \\ 
  C2053*   &    J100102.27+023034.5   &   82     &   0   &  150.259458    &   2.509566     &  8.0       &     41    &         9     &      66   &      10   \\   
  C2078*   &    J100104.26+023307.7   &   79     &   0  &   150.267716     &  2.552051     &  6.6      &      44     &        8    &       46    &      8         \\
  C2096*   &    J100105.55+023309.8   &   158     &    1   &  150.273142  &  2.55273 &  6.3  & - & - & 77 &  11 \\ 
  C2096a  &   J100105.55+023309.8   &   158     &   0   &  150.273147   &    2.552731    &   6.7          &  37     &        8     &      39    &      8        \\   
  C2096b   &   J100105.55+023309.8   &   158      &    0   &  150.273136    &   2.552728   &    6.6          &  30     &        7      &     38       &   8        \\
  C2136   &   J100108.99+022815.9   &   669     &  0  &   150.287493   &    2.471066  &     10.3          &  389       &      40      &     390    &     40        \\ 
  C2184   &   J100112.06+024106.7   &   2464      &   1   &  150.300291 & 2.685168 & 8.7 & - & -  & 611  & 47       \\   
  C2184a  &   J100112.06+024106.7   &   2464       &  0   &  150.300296    &   2.685171   &    9.5          &  158        &     19   &        221   &        24       \\
  C2184b  &   J100112.06+024106.7   &   2464     &   0   &  150.300288    &   2.685166   &    9.4           &  143        &     17     &      390     &     40        \\ 
  C2231*  &    J100116.17+023241.8   &   133     &   0   &  150.317413     &  2.544883     &  5.3         &   26          &   6   &        33       &    6         \\   
  C2259   &    J100118.04+023610.3   &   77   &   0  &   150.325164    &   2.602793   &    5.6          &  60      &       8    &       67    &      9       \\
  C2283   &    J100119.92+023856.2   &   326    &     0   &  150.332979     &  2.648923   &    6.6          &  38         &    8    &         43       &   8     \\ 
  C2284   &   J100120.06+023443.7   &   10590      &     0   &  150.3336   &    2.578797    &   13.2          &  8860        &     891     &      9215      &    922     \\   
  C2290   &    J100120.45+023834.2   &   183    &    0   &  150.335177   &    2.642863   &    6.3           &  41          &   8    &       45   &       8      \\
  C2302  &  J100121.29+023535.8   &   228     &    0  &   150.33873    &   2.593252   &    5.3          &  42        &     7   &        57     &     8       \\ 
  C2313   &   J100122.02+023724.3   &   224    &     0   &  150.341768    &   2.623411    &   5.8          &  139          &   15     &      140     &     15     \\   
  C2315*  &   J100122.07+023405.5   &   235     &   0  &   150.341957    &   2.568208     &  5.2      &      36     &       6    &       82     &     10         \\
  C2361*  &   J100125.31+023527.5   &   131     &   0  &   150.355530    &   2.590917    &   5.2        &     20   &         6      &     27     &     6         \\ 
  C2362  &   J100125.36+023851.7   &   134     &   0  &   150.355690    &   2.647704  &     6.1          &  78        &     10     &      80       &   10       \\   
  C2374* &   J100126.28+023934.1   &   57     &  0  &   150.359571    &   2.659473   &    7.6        &    39     &        9  &         42     &    9           \\
  C2383*  &   J100127.97+024029.3   &   1502     &  0   &  150.366624   &    2.674827   &    8.1       &     42      &       9      &      57     &     10          \\   
  C2407  &   J100129.83+023239.0   &   159     &  0   &  150.374252    &   2.544137   &    6.1          &  63      &       9     &      76       &   10        \\
  C2436   &   J100131.14+022924.7   &   5377    &    0    & 150.379782     &  2.490183   &    9.9          &  923     &        93    &       1184   &       119      \\ 
  C2443   &  J100131.36+022639.2   &   16120     &     0  &   150.380641   &    2.444316  &     11.4          &  70      &       13       &    83    &       14    \\   
  C2470   &   J100133.58+022749.6   &   127    &    0   &  150.389932     &  2.463782    &   10.1          &  59       &      12   &        60    &      12      \\
  C2512   &   J100136.93+023834.3   &   264   &    0  &   150.403918   &    2.642866   &    7.6          &  166       &      18       &    173    &      19      \\ 
  C2517   &   J100137.85+023710.4   &   80    &    0   &  150.407764     &  2.619514   &    7.2          &  56           &  9      &     56   &       9      \\   
  C2535  &  J100139.36+023432.0  &    154     &      0   &  150.41403   &    2.575519  &     7.2          &  68        &     10    &       79  &       11     \\
  C2541*   &   J100139.85+023329.3   &   166     &    0  &   150.416043    &   2.558077     &  7.3        &     39        &     8     &      43     &     8        \\ 
  C2566*   &   J100141.42+023523.9   &   84     &  0   &  150.422577   &    2.589912   &    7.2        &    48      &       9        &   50 &         9          \\   
  C2627   &    J100145.97+023948.0   &   262   &     0   &  150.441538    &   2.663279  &     11.8          &  110      &       16   &        166      &    20     \\
  C2631   &    J100146.69+023251.6   &   122    &     0   &  150.444606   &    2.547659    &   9.2          &  77       &      12      &     86     &     13     \\ 
  C2662   &   J100149.61+023334.8   &   2202      &     1    &  150.456723 & 2.559672  &  10.3 &  - & -  & 1178  & 96  \\   
  C2662a   &   J100149.61+023334.8   &   2202      &    0   &  150.456724    &   2.559673   &    13.3          &  838         &    85    &       891   &       90      \\
  C2662b  &  J100149.61+023334.8   &   2202      &   0  &   150.456718   &    2.55967   &    13.1          &  232      &      27       &    287      &     32       \\ 
\hline
\end{tabular}
\begin{tablenotes}
%\small
\item {\textbf{Notes.}} {\it {Column 1}}: Source name used in the present project. An asterisk (*) represents those sources that were detected with the VLBA+GBT but not with the VLBA. A lower case letter added to the ID refers to each component of a multi-component source; {\it {Col. 2}}: Source name from \citet{schinnerer2010}; {\it {Col. 3}}: Integrated VLA flux density of the source (1.4 GHz) from \citet{schinnerer2010}; {\it {Col. 4}}: Classification between single- and multi-component source, 0: single-component source, 1: multi-component source; {\it {Cols. 5, 6}}: Right ascension and declination (J2000) of the source, measured with the VLBA+GBT (uniform weighting); {\it {Col. 7}}: Local noise rms measured with the VLBA+GBT (uniform weighting); {\it {Cols. 8, 9}}: Peak flux density of the source and its error, measured with the VLBA+GBT (uniform weighting); {\it {Cols. 10, 11}}: Integrated flux density of the source and its error, measured with the VLBA+GBT (uniform weighting).
\end{tablenotes}
\end{threeparttable}
\end{table*}

\begin{table*}
\centering 
\caption{Multi-wavelength information of the COSMOS VLBA+GBT detected sources previously undetected by the VLBA alone.}             
\label{table:mwgbt}     
\begin{threeparttable}                       
\begin{tabular}{| l | l | l | l | l | l | l | l | l |}      
\hline\hline           
ID & zphot & zspec & S3.6um & S24um & S\_sX(0.5-2 keV) & S\_hX(2-10 keV) & Mph & logM* \\    
  &   &    & [$\mu$Jy]  & [$\mu$Jy] & [10$^{-7}$\,W/cm$^{2}$] & [10$^{-7}$\,W/cm$^{2}$] &  & [log(M$_\odot$)]  \\
  (1) & (2) & (3) & (4) & (5) & (6) & (7) & (8) & (9)  \\
\hline                        
  C2053  &  0.73  &  0.902 &   27.12  &  30.4      &       &       &  1 &  10.6   \\
  C2078  &  0.22  &  0.267 &  252.4  &  169.2   &  7.7$\cdot$10$^{-16}$ &  4.03$\cdot$10$^{-15}$  &  1 &   10.93      \\
  C2096  &  0.69  &        &   85.43  &  119.0 & 4.7$\cdot$10$^{-16}$  &  2.15$\cdot$10$^{-15}$     &  2 &   11.15      \\
  C2231  &        &        &  2.06    &  18.8 &       &       &    &         \\
  C2315  &  0.81  &  0.821 &   33.97  &  186.0 &       &       &  2 &   10.4 \\
  C2361  &  1.99  &        &   23.84  &  545.7 &       &       &  3 &   10.84 \\
  C2374  &  0.4   &  0.431 &   102.3  &  34.4  &       &       &  1 &   11.03 \\
  C2383  &  1.09  &  0.68  &   62.63  &        &       &       &  1 &   11.18 \\
  C2541  &  1.17  &        &    35.3  &  96.8  &       &       &  2 &   10.68 \\
  C2566  &  0.76  &        &    44.9  &  108.2 &       &       &  2 &    10.81 \\
\hline                                  
\end{tabular}
\begin{tablenotes}
%\small
\item {\textbf{Notes.}} {\it {Column (1)}}: Source name used in the present project; {\it {Column (2)}}: Photometric redshift from \citet{laigle2016}; {\it {Column (3)}}: Spectroscopic redshift from \citet{lilly2007} and \citet{gabor2009} (C2383); {\it {Column (4)}}: Spitzer/IRAC 3.6 $\mu$m flux density from the Spitzer Enhanced Imaging Products (SEIP) source list from the NASA/IPAC Infrared Science Archive\protect\footnotemark, in $\mu$Jy; {\it {Column (5)}}: Spitzer/MIPS 24 $\mu$m flux density from \citet{muzzin2013}, in $\mu$Jy; {\it {Columns (6) and (7)}}: Soft (0.5-2 keV) and hard (2-10 keV) band X-ray fluxes from \citet{civano2016}, in 10$^{-7}$\,W/cm$^{2}$; {\it {Column (8)}}: Morphological classification from \citet{tasca2009}, 1: early type, 2: spiral, 3: irregular; {\it {Column (9)}}: Stellar mass of the galaxy from \citet{laigle2016}, in log(M$_\odot$).
\end{tablenotes}
\end{threeparttable}
\end{table*}

\begin{figure*}[t!]
\centering
\includegraphics[width=\textwidth, height=6.5cm]{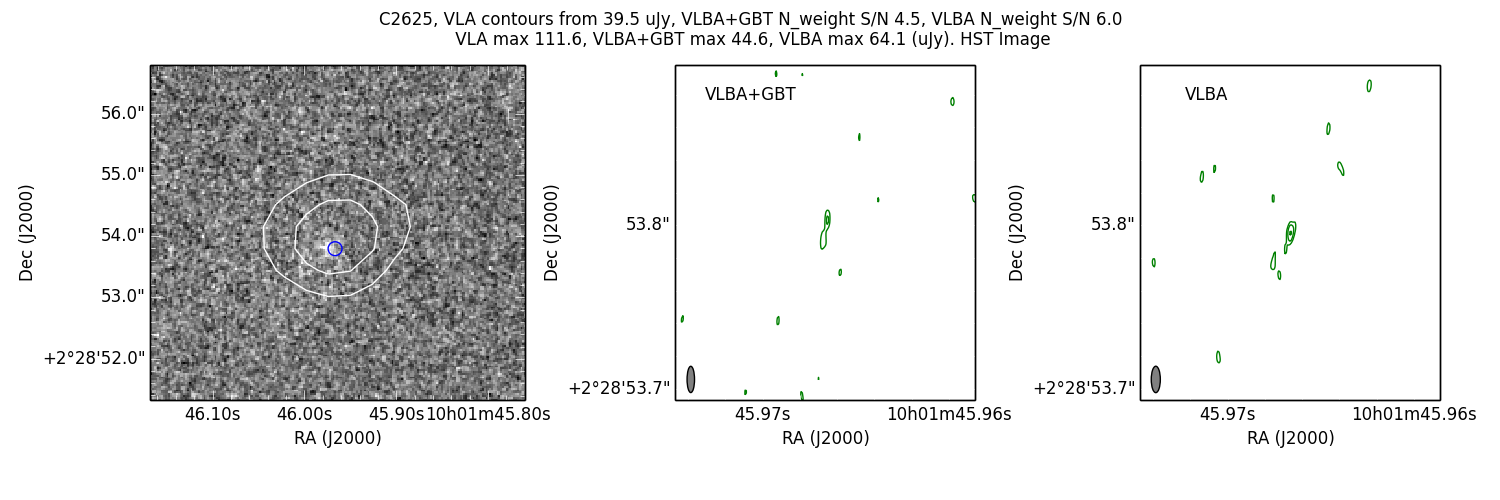}
\caption{Optical counterpart and radio contour plots of C2625. The header contains: i) The source name used in the present project; ii) the rms noise value at which the VLA contours start; iii) The VLBA+GBT naturally-weighted image S/N; iv) The VLBA naturally-weighted image S/N; v) The VLA peak flux density (in $\mu$Jy); vi) The VLBA+GBT peak flux density (in $\mu$Jy); vii)  The VLBA peak flux density (in $\mu$Jy). {\em {Left panel:}} The background greyscale image is the HST image of the optical counterpart \citep{koekemoer2007, massey2010}. The white contours represent the VLA contours of the source, starting at four times the rms noise level of the VLA image and increasing by a factor of two. The blue circle represents the VLBA detection position. {\em {Middle panel:}} Green contours represent the VLBA+GBT contours of the source, starting at three times the rms noise level of the naturally-weighted image and increasing by a factor of $\sqrt{2}$. {\em {Right panel:}} Green contours represent the VLBA detection contours, starting at three times the rms noise level of the naturally-weighted image and increasing by a factor of $\sqrt{2}$.}
\label{fig:count}
\end{figure*}

We followed the same methods for source extraction and to measure the VLBA+GBT flux density and position of the sources as those explained in detail in \citet{herreraruiz2017}. Once all the VLBA+GBT data were calibrated and imaged, we found 35 detected sources with a signal-to-noise ratio (S/N) greater than 5.5. Three of them were found to be multi-component sources. Ten of them were not detected before with the VLBA (because they would have been below the detection threshold of the VLBA observations, i.e., 55\,$\mu$Jy). One source of the input sample (C2625) was detected in the 2012--2013 observations by the VLBA alone, but was not detected with S/N\,>\,5.5 in our new VLBA+GBT data set (although a source is seen at lower significance). Figure~\ref{fig:count} shows the optical counterpart\footnote{\url{http://irsa.ipac.caltech.edu/data/COSMOS/index_cutouts.html}} and the VLBA+GBT and the VLBA contours of the C2625 images. One possible explanation is that C2625 might have been in a high flux density state in the moment of the VLBA observations and in a lower state in the moment of the VLBA+GBT observations, since this source is among the faintest VLBA detected sources. We reprocessed the image of C2625 excluding all baselines involving the GBT in order to create a VLBA-only image. With this exercise, we wanted to test the possibility of the GBT baselines being the reason of the non-detection by, e.g., introducing a different uv sampling. However, the result was similar to the one obtained including the GBT baselines, which supports the conclusion that the source is indeed variable, with a decline in flux density by $\sim$30\% over a period of 3\,years.

Table~\ref{table:agncatgbt} is the catalogue of the 35 VLBA+GBT detected sources. In the case of multi-component sources, we have added a lower case letter to the source ID for each component and we have included a new line in the catalogue, which contains the original ID (i.e., no lower case letter added) and the total integrated flux density that corresponds to the sum of the integrated flux density of each component.

We collected additional multiwavelength information for the 10 newly VLBA+GBT detected sources (see Table~\ref{table:mwgbt}). The multiwavelength information for the rest of the sample, i.e., previously detected sources with the VLBA, can be found in \citet{herreraruiz2017}. Counterparts have been associated via nearest neighbour matching (1\,arcsec radius). 

As we can see in Table~\ref{table:mwgbt}, we find a rather low number of X-ray counterparts for the VLBA+GBT detected sources. This result has been amply discussed in \citet{herreraruiz2017}, where we compared the sample of the VLBA detected sources to the sample of low-to-moderate radiative luminosity AGN (MLAGN) from \citet{smolcic2017} and found that most of the VLBA detected sources without an X-ray counterpart were also classified as MLAGN. We concluded then that X-ray surveys might miss AGN with low accretion rates, i.e., radiatively inefficient AGN. Heavily obscured, Compton-thick ($N_\mathrm{H} > 10^{24} $\,cm$^{-2}$) AGN might as well being missed by X-ray surveys (e.g., \citealt{xue2017}). However, \citet{lanzuisi2015} analysed heavily obscured AGN in the XMM-COSMOS survey and obtained a final sample of ten Compton-thick AGN, none of which correspond to our VLBA+GBT sample.

\footnotetext{\url{http://irsa.ipac.caltech.edu/Missions/spitzer.html}}

\section{Euclidean-normalized differential source counts}
\label{sec:res}

Differential radio source counts, d$N$/d$S$, have been frequently used to measure the contribution of AGN to the faint radio population. They represent the number of sources per flux density interval per unit solid angle as a function of flux density:

\begin{equation}
\dfrac{\mathrm{d}N}{\mathrm{d}S} = \dfrac{N}{\Omega \Delta S}
\end{equation} 

\noindent where {\it {N}} is the number of sources in a flux density bin, $\Omega$ is related to the area over which a source with flux density $\langle S \rangle$ (mean flux density of the bin) could be detected, and $\Delta S$ is the width of the flux density bin.

We have analysed the Euclidean-normalised differential radio source counts in the $\mu$Jy regime to estimate a lower limit for the AGN contribution to the faint radio population. For this purpose, we have used the sample of the VLBA detected sources \citep{herreraruiz2017} with the addition of the new sources detected with the VLBA+GBT. This sample is hereafter referred to as our VLBI sample. Our VLBI sample constitutes an AGN sample since only their brightness temperatures are high enough (>10$^{5}$\,K) for a VLBI detection (see \citealt{condon1992} for a detailed review of radio emission from galaxies). Occasionally, star-forming activity and transient events like radio supernovae and gamma-ray bursts can also reach this limit. Nevertheless, the transient events are quite rare at radio wavelengths and the luminosity of star formation drops below our detection threshold when the redshift of the host galaxy is larger than 0.1, where almost all of our sources are located. Moreover, the luminosities of our VLBI detected sources are all larger than 10$^{21}$\,W/Hz, i.e., above the limit chosen by \citet{kewley2000} to separate AGN from star-forming galaxies. Our AGN selection technique (based only on the large brightness temperatures of the sources) is then independent of the multiwavelength properties of the source, what gives us the opportunity to compare our results to those from other deep radio surveys that rely on them. 

As an example, the $q_{24}$ parameter (ratio between the measured 24\,$\mu$m flux density and the measured 1.4\,GHz flux density, $q_{24}$=log($S_{24\mathrm{\mu m}}/S_{\mathrm{1.4GHz}}$)) is one of the observables generally used by radio surveys using multiwavelength diagnostics to classify their sources between radio-loud (RL) AGN, radio-quiet (RQ) AGN and star-forming galaxies. \citet{ibar2008} analysed the 24\,$\mu$m properties of a radio-selected sample and characterized the transition from RL AGN to RQ AGN and star-forming galaxies at faint ($\lesssim$1\,mJy) flux densities. They classified a source as a RL if $q_{24} < -0.23$. As we have previously discussed, our VLBI observations are sensitive to AGN, and so the $q_{24}$ parameter would separate our VLBI detected sources between RL AGN and RQ AGN. Taking the threshold from \citet{ibar2008}, we found that 66\% of the VLBI detected sources (which have available 24\,$\mu$m measurements) would be classified as RQ AGN and 34\% as RL AGN.

The amount of detected sources in this project represents a lower limit on the number of AGN in the area. This is important in particular at faint flux densities, where the sensitivity of the observations and the compactness of the source play an important role. It is worth noting that we have used the VLA flux densities of our VLBI sample to construct the radio source counts because the high resolution of VLBI observations would resolve out a large percentage of their flux densities. Although bright radio AGN are expected to be highly variable, this might not be necessarily the case for the faint radio population. \citet{carilli2003} analysed the variability of radio sources in the Lockman Hole region and found that the radio sky is not highly variable at the sub-mJy level. Moreover, variability seems to be more important at higher frequencies (e.g., \citealt{aller1985, mooley2016}). Therefore, we do not expect a significant effect on the radio source counts due to variability. 

% They estimated the areal density of highly variable sources to be $<$5$\times$10$^{-3}$\,arcmin$^{-2}$ at that level. 

Two effects need to be taken into account in order to derive the radio source counts, concerning i) the estimation of the effective area (area where a source would have been detected), where the sensitivity of the observations plays an important role, and ii) the completeness of the survey, affected by the resolution of the observations. Extended sources are also affected by the resolution since part of their total radio flux density can be resolved out.

To ensure the right implementation of the procedure to derive the radio source counts, we first used the VLA sample from the VLA-COSMOS Large Project Catalog \citep{schinnerer2007} to construct the radio source counts. We then compared our results to those obtained by \citet{bondi2008}, who analysed the radio source counts using the same VLA sample. The radio source counts using the VLA sample from \citet{schinnerer2010}, which was our input catalogue, have not been analysed. Therefore, not all the sources in our VLBI sample might have been contained in the catalogue of \citet{schinnerer2007}. We cross-matched our VLBI sample with the catalogue of \citet{schinnerer2007} and we found 316 counterparts in the inner 1\,deg$^{2}$ of the COSMOS field (1\,arcsec matching radius), which was the region where the sensitivity was the deepest and mostly uniform.

\begin{figure}
\centering
\includegraphics[scale=0.49]{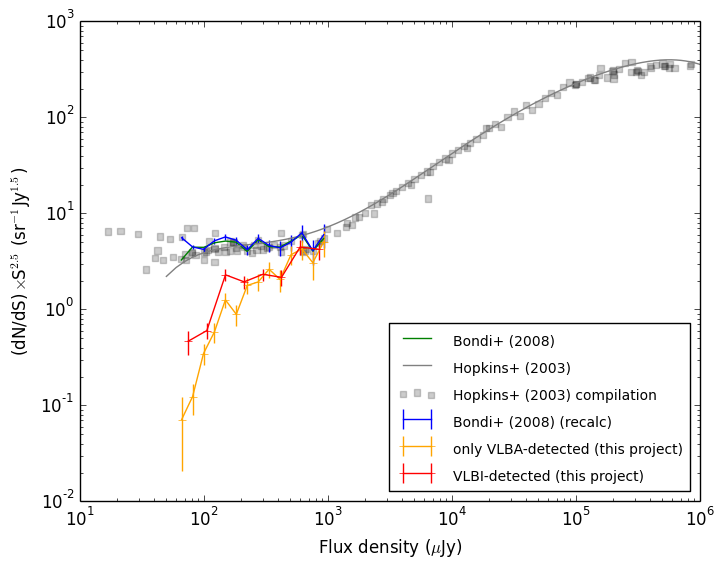}
\caption{The 1.4 GHz Euclidean-normalised source counts (lower limits) of the VLBI-detected sources (red line) of the COSMOS field. We have used the associated VLA flux densities of the VLBI detected sources. The orange line represents the derived source counts (lower limits) using only the sample of VLBA detected sources. The green line represents the source counts from \citet{bondi2008} and the blue line is the reconstruction of their results using the same procedure as for the VLBI-detected sources. The grey squares represent the data from \citet{hopkins2003} and the grey line their polynomial fit.}
\label{fig:scou}
\end{figure}

In Fig.~\ref{fig:scou}, we show the reconstructed radio source counts following the procedure described by \citet{bondi2008} (blue line) and the one published by \citet{bondi2008} (green line). One can see that both results are in very good agreement with the exception of the lowest flux density bin. A small variation of the procedure to select the sources of the sample used by \citet{bondi2008}, in particular the faintest ones, might be a possible reason for this difference.

We used then our VLBI sample to calculate the radio source counts using the same procedure. In this case, the effective area is a combination of the areas where a source would have been detected with the VLA and with the VLBA/VLBA+GBT. We determined the VLBA and VLBA+GBT effective areas using their sensitivity maps (see \citealt{herreraruiz2017} for the VLBA and Sect.~\ref{sec:sensmapgbt} of this paper for the VLBA+GBT) and considering a S/N threshold of 5.5. It is worth noting that by using the VLBI sensitivity map to derive the effective area, we are implicitly assuming that 100\% of the source flux density is contained in the VLBI-detectable compact component. As we showed in \citet{herreraruiz2017}, this is frequently not the case. Furthermore, \citet{rees2016} have shown, using our VLBA data set, that only about 50\% of radio-loud AGN are detected by VLBI, and this fraction is
independent of the properties of the host galaxy, and does not depend strongly on radio flux density. Therefore, particularly near the sensitivity limit, we will underestimate the AGN counts and so we derive a lower limit for the AGN contribution.

Figure~\ref{fig:scou} shows the Euclidean-normalised radio source counts (lower limits) constructed with our VLBI sample (using the associated VLA flux densities). We present in Table~\ref{table:scountv} the flux density bin, $S_{\mathrm{bin}}$, the mean flux density, $\langle S \rangle$, the number of sources observed, $N$, the lower limits for the differential source counts, d$N$/d$S$, and the lower limits for the Euclidean-normalised differential source counts, (d$N$/d$S$)$S^{2.5}$, with their estimated Poissonian errors. In addition, we have plotted in Fig.~\ref{fig:scou} the derived source counts considering only the sample of VLBA detected sources, i.e., not taking into account the sample of VLBA+GBT detected sources. We can see that with the addition of the GBT to the VLBA we are able to recover a higher number of faint sources.

\begin{table}
\caption{The 1.4 GHz radio source counts (lower limits) of the VLBI-detected sources of the COSMOS field} 
\label{table:scountv}     
\renewcommand{\arraystretch}{1.25}
\centering                         
\begin{tabular}{c c c c c}        
\hline\hline
$S_{\mathrm{bin}}$ & $\langle S \rangle$ & $N$ & d$N$/d$S$ & (d$N$/d$S$)$S^{2.5}$ \\   
$\mu$Jy & $\mu$Jy &  & (sr$^{-1}$Jy$^{-1}$) &  (sr$^{-1}$Jy$^{1.5}$) \\   
\hline   
63.0-89.1     &   74.9  &   12   &  >9.59$\cdot 10^{9}$  &   >0.466$\pm$0.135 \\   
89.1-126.0    &  106.0  &   38   &  >5.21$\cdot 10^{9}$  &   >0.602$\pm$0.112 \\   
126.0-178.2   &  149.8  &   66   &  >8.34$\cdot 10^{9}$  &   >2.292$\pm$0.338 \\   
178.2-252.0   &  211.9  &   49   &  >2.96$\cdot 10^{9}$  &   >1.936$\pm$0.292 \\   
252.0-356.4   &  299.7  &   47   &  >1.49$\cdot 10^{9}$  &   >2.320$\pm$0.338 \\   
356.4-504.0   &  423.8  &   26   &  >5.84$\cdot 10^{8}$  &   >2.159$\pm$0.423 \\   
504.0-712.8   &  599.4  &   32   &  >5.08$\cdot 10^{8}$  &   >4.468$\pm$0.790 \\   
712.8-1008.0  &  847.6  &   18   &  >2.02$\cdot 10^{8}$  &   >4.228$\pm$0.997 \\   
\hline                               
\end{tabular}
\end{table}

A similar flattening to the total source counts seems to appear also in the case of the source counts of the VLBI-detected sources at flux densities between $\sim$100-500\,$\mu$Jy. The AGN contribution to the total source counts at that flux density range accounts for >40-55\%. The AGN contribution is larger ($>$75\%) at higher flux densities ($\sim$0.5-1\,mJy) and it may drop further at low flux densities ($<$100\,$\mu$Jy), although our results there are only relatively weak lower limits, as discussed in the following section. Additionally, we calculated the ratio between the measured flux densities using VLBI and the VLA flux densities of our VLBI sample and found the median value to be 0.63. This means that at least 63\% of the radio emission comes from the AGN, since the VLBI observations provide a lower limit on the AGN emission. The difference of the radio emission might come either from star formation, since it would have been resolved out by VLBI, or from more extended radio emission of the AGN, e.g., large-scale jets and lobes. Then, the star formation contribution here should be considered only as an upper limit. This is in agreement with the results from \citet{ballantyne2009}, who argued that the $\mu$Jy AGN population might undergo moderate levels of star formation, making them ideal objects for future studies of the AGN-host galaxy co-evolution.

Our results strengthen the hypothesis that a significant fraction of the faint (sub-mJy) radio population is AGN-powered, rather than being composed solely of star-forming galaxies, especially since these fractions represent a lower limit.

\section{Discussion}
\label{sec:dis}

\begin{figure}
\centering
\includegraphics[scale=0.48]{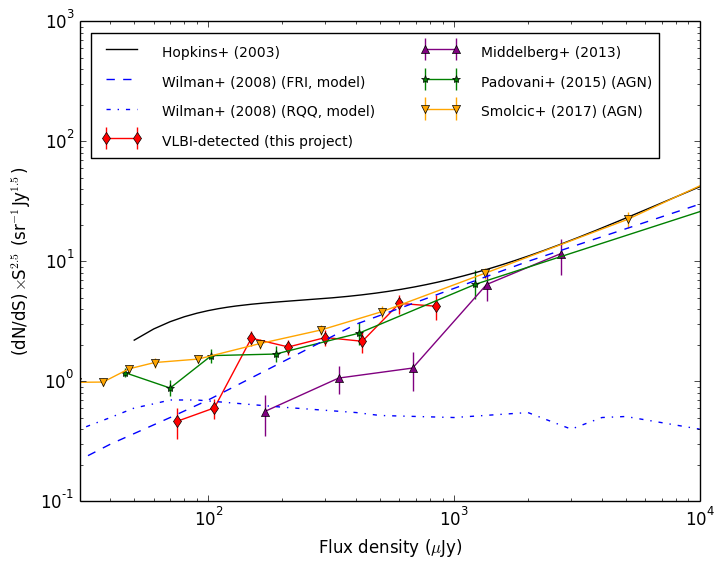}
\caption{Euclidean-normalised source counts comparison to the literature. The black line represents the polynomial fit to the total source counts from \citet{hopkins2003}. The dashed and dot-dashed blue lines represent the predicted source counts of Fanaroff-Riley class I (FRI) radio galaxies and radio-quiet quasars (RQQ), respectively, by \citet{wilman2008}. The red diamonds represent the lower limits for the source counts of our VLBI-detected sources. The purple triangles represent the source counts of the VLBI-detected sources by \citet{middelberg2013}. The green stars represent the AGN source counts from \citet{padovani2015}. The orange inverted triangles represent the AGN source counts from \citet{smolcic2017}.}
\label{fig:scoucomp}
\end{figure}

We compare our derived source counts with several projects with independent estimates available in the literature (see Fig.~\ref{fig:scoucomp}). In particular, \citet{padovani2015} and \citet{smolcic2017} analysed the radio source counts using completely different methods of AGN identification. \citet{padovani2015} studied the Extended Chandra Deep Field South (ECDFS) VLA sample at 1.4\,GHz. Their AGN classification was based either on a radio excess shown by the {\it{q}}$_{24obs}$ parameter (the ratio between the observed 24\,$\mu$m and 1.4\,GHz flux densities) or on evidences of AGN activity in other bands (hard X-rays or IRAC colour-colour diagram). \citet{smolcic2017} studied the faint radio population composition from the VLA-COSMOS 3\,GHz Large Project \citep{smolcic2017b}. Their AGN classification was based on various criteria involving X-ray luminosity, observed mid-infrared colour, ultraviolet-far-infrared spectral energy distribution, rest-frame near-ultraviolet-optical colour corrected for dust extinction, and radio-excess relative to that expected from the star-formation rate of the hosts (see also \citealt{delvecchio2017}). At flux densities between $\sim$150\,$\mu$Jy and 1\,mJy, our derived source counts of VLBI-detected sources are in very good agreement with the results from \citet{padovani2015} and \citet{smolcic2017}. 

\citet{wilman2008} performed simulations of the extragalactic radio continuum sky to optimize the design of future radio interferometers like the Square Kilometre Array (SKA). The VLBI observations should be able to detect the Fanaroff-Riley class I (FRI) radio galaxies and the radio-quiet quasar populations. Therefore, our derived source counts should be compared to the predicted FRI+RQQ source counts by \citet{wilman2008}, which would lie close to the observational results from \citet{padovani2015} and \citet{smolcic2017}. They are in good agreement except for the faint end of our derived source counts, which seems to follow only the FRI population. We discuss this drop in the following.

As mentioned before, adding the GBT to the VLBA to increase sensitivity leads to a higher detection fraction of faint sources. Moreover, our derived source counts represent a lower limit on the AGN contribution to the radio population since we have used the VLBI sensitivity maps to estimate the effective area. Therefore, the sensitivity limit of our VLBI observations is most probably the reason for the drop of our derived source counts at flux densities below 100\,$\mu$Jy. If we consider the flux density range between 180 and 360\,$\mu$Jy, we find that 50\% of the VLBA+GBT detected sources have a VLBI brightness lower than 50\% of the VLA flux density. In the case of flux densities below 100\,$\mu$Jy, we find that none of the VLBA+GBT detected sources have a VLBI brightness below this threshold, which means that near the sensitivity limit only sources that are almost entirely compact on milliarcsecond scales can be detected. \citet{middelberg2013} analysed the radio source counts using a sample of VLBI-detected sources in the Lockman Hole/XMM field. Their observations had a 1\,$\sigma$ sensitivity of $\sim$20\,$\mu$Jy\,beam$^{-1}$ near the pointing centre and they used a 6\,$\sigma$ detection threshold. Our observations are therefore $\sim$7\,\,times more sensitive and we can see in Fig.~\ref{fig:scoucomp} that the drop of our derived source counts occurs at a flux density $\sim$7\,\,times lower than the flux density at which the drop of the source counts from \citet{middelberg2013} occurs, which is consistent with the non-detection of partially resolved sources being the main culprit for the drop-out in both cases. \citet{middelberg2013} also carried out an additional test to demonstrate that the sensitivity of the observations plays an important role in the construction of the radio source counts. They suggested that the source counts of VLBI detected sources would rise towards the total one with better sensitivities. This is exactly what we see using our VLBI sample, whose VLBI observations had a better sensitivity than the one by \citet{middelberg2013} (see Fig.~\ref{fig:scoucomp}).

Completely different observational methods of AGN identification produce very similar results, making the AGN contribution to the sub-mJy radio population strongly reliable.

\section{Conclusions}
\label{sec:con}

We have observed one pointing with 179 sources of the COSMOS field using the VLBA+GBT at 1.4\,GHz. The main goal of the observations was to achieve a better sensitivity than that obtained with the VLBA observations \citep{herreraruiz2017} in order to reach even fainter sources. We have analysed the primary beam response of the GBT to calibrate the VLBA+GBT data as it was yet untested in VLBI observations. We have calibrated and analysed the data, obtaining a $1\sigma$ rms noise level of $\sim$3\,$\mu$Jy in the central part of the pointing and milli-arcsecond resolution images. The following are the main results of this project:

   \begin{enumerate}
      \item We have detected 35 sources in a single pointing with the VLBA+GBT, 10 of these were previously not detected using the VLBA only. We have constructed and presented the catalogue of the VLBA+GBT detected sources along with a catalogue containing multiwavelength information of the 10 newly detected sources. 
      \item The VLBA+GBT detected sources can be considered as AGN because of the very high brightness temperature (>\,10$^{5}$\,K) that an object needs to have to be detected with VLBI observations (when the host galaxy of the object is located at a redshift larger than 0.1, where all the VLBA+GBT detected sources are). 
      \item We have constructed the Euclidean-normalised radio source counts using the sample of VLBA detected sources \citep{herreraruiz2017} together with the sample of the new sources detected with the VLBA+GBT. The obtained radio source counts represent a lower limit on the AGN fraction of the faint radio population. At flux densities between $\sim$100-500\,$\mu$Jy, we found a lower limit for the AGN contribution of >40-55\% to the total source counts. This percentage increases for higher flux densities. Our results are in very good agreement with the findings from \citet{padovani2015} and \citet{smolcic2017}, which use entirely different methods of AGN identification. At flux densities below 100\,$\mu$Jy, the source counts of our VLBI detected sources exhibit a steep drop. We think that the sensitivity limit of our VLBI observations is the reason for this drop. We see no evidence for a change in the AGN fraction at $\sim$100\,$\mu$Jy, implying that the expected eventual drop-out occurs at fainter flux densities.

      \item Very similar results are produced by different observational lines of evidence, which strengthens the conclusion that a significant fraction of the sub-mJy radio population are radio-emitting AGN.
   \end{enumerate}

\vspace{0.5cm}

\begin{acknowledgements}
      N.H.R. acknowledges support from the Deutsche Forschungsgemeinschaft through project MI 1230/4-1. V.S., M.N. and I.D. acknowledge the European Union’s Seventh Framework programme under grant agreement 337595 (ERC Starting Grant, ``CoSMass''). P.N.B. is grateful for support from STFC via grant ST/M001229/1. We thank the anonymous referee for the helpful and constructive comments, which have improved this paper. This work made use of the Swinburne University of Technology software correlator, developed as part of the Australian Major National Research Facilities Programme and operated under licence. This work made use of \texttt{Topcat} \citep{taylor2005}, available at \url{http://www.starlink.ac.uk/topcat/}. This work also made use of \texttt{APLpy}, an open-source plotting package for Python hosted at \url{http://aplpy.github.com}, and  \texttt{Astropy}, a community-developed core Python package for Astronomy \citep{astropy2013}. We wish to thank the staff of the VLBA and the GBT who greatly supported the experimental observations in this project. The VLBA and the GBT are operated by the Long Baseline Observatory and the Green Bank Observatory, respectively, facilities of the National Science Foundation operated under cooperative agreement by Associated Universities, Inc.
\end{acknowledgements}

\setlength{\bibsep}{0pt plus 0.3ex}
\bibliographystyle{agsm}
\bibliography{references}

%
% Online Material
%_____________________________________________________________
%        Online appendices have to be placed at the end, after
%                                        \end{thebibliography}
%-------------------------------------------------------------
%\end{thebibliography}

%\Online
\newpage

\clearpage
\onecolumn

\begin{appendix} %First online appendix
\section{List of the targeted sources}
\label{sec:appx}

\begin{longtab}
\begin{longtable}{llllll}
\caption{\label{targetcat} List of the COSMOS VLBA+GBT targeted sources. {\it {Col. 1}}: Source name used in the present project; {\it {Col. 2}}: Source name from \citet{schinnerer2010}; {\it {Col. 3}}: Integrated VLA flux density of the source (1.4 GHz) from \citet{schinnerer2010}; {\it {Col. 4}}: Classification between single- and multi-component source from \citet{schinnerer2010}, 0: single-component source, 1: multi-component source; {\it {Cols. 5, 6}}: Right ascension and declination (J2000) of the source from \citet{schinnerer2010}.}\\
\hline\hline
ID & COSMOSVLADP & S$_{i,VLA}$ ($\mu$Jy) & M$_{VLA}$ & RA$_{VLA}$ (deg) & Dec$_{VLA}$ (deg)  \\
  (1) & (2) & (3) & (4) & (5) & (6)  \\
\hline
\endfirsthead
\caption{continued.}\\
\hline\hline
ID & COSMOSVLADP & S$_{i,VLA}$ ($\mu$Jy)  & M$_{VLA}$ & RA$_{VLA}$ (deg) & Dec$_{VLA}$ (deg)  \\
  (1) & (2) & (3) & (4) & (5) & (6)  \\
\hline
\endhead
\hline
\endfoot
  C1826  &  J100044.74+023326.5  &    75  & 0 &  150.186450 &  2.557375  \\
  C1881  &  J100048.06+023441.6  &    71  & 0 &  150.200254 &  2.578244  \\
  C1886  &  J100048.89+023127.5  &   234  & 0 &  150.203733 &  2.524325  \\
  C1903  &  J100050.45+023356.1  &   610  & 0 &  150.210229 &  2.565591  \\
  C1904  &  J100050.49+023254.5  &    79  & 0 &  150.210412 &  2.548488  \\
  C1908  &  J100050.84+023817.6  &   234  & 0 &  150.211866 &  2.638227  \\
  C1914  &  J100051.50+022918.5  &    62  & 0 &  150.214583 &  2.488477  \\
  C1915  &  J100051.58+023334.2  &   122  & 0 &  150.214941 &  2.559511  \\
  C1916  &  J100051.97+023529.2  &   100  & 0 &  150.216541 &  2.591447  \\
  C1934  &  J100054.47+023409.2  &    83  & 0 &  150.226991 &  2.569244  \\
  C1935  &  J100054.49+023905.1  &    73  & 0 &  150.227050 &  2.651438  \\
  C1936  &  J100054.50+023215.4  &   213  & 0 &  150.227083 &  2.537611  \\
  C1942  &  J100054.83+023126.1  &   201  & 0 &  150.228470 &  2.523927  \\
  C1944  &  J100054.87+023314.1  &    76  & 0 &  150.228624 &  2.553933  \\
  C1946  &  J100054.99+022849.9  &    92  & 0 &  150.229162 &  2.480530  \\
  C1948  &  J100055.31+022942.0  &   196  & 0 &  150.230458 &  2.495005  \\
  C1950  &  J100055.37+023441.5  &    78  & 0 &  150.230741 &  2.578208  \\
  C1952  &  J100055.73+023528.7  &    61  & 0 &  150.232245 &  2.591308  \\
  C1953  &  J100055.79+023705.8  &    88  & 0 &  150.232458 &  2.618283  \\
  C1960  &  J100056.12+023939.0  &    67  & 0 &  150.233837 &  2.660836  \\
  C1965  &  J100056.43+023812.5  &    78  & 0 &  150.235124 &  2.636808  \\
  C1969  &  J100056.77+023841.5  &    45  & 0 &  150.236566 &  2.644861  \\
  C1972  &  J100056.94+024120.9  &   105  & 0 &  150.237283 &  2.689147  \\
  C1975  &  J100057.06+022942.9  &   123  & 0 &  150.237766 &  2.495275  \\
  C1977  &  J100057.11+023451.7  &   347  & 0 &  150.237958 &  2.581044  \\
  C1985  &  J100057.44+023620.6  &   102  & 0 &  150.239350 &  2.605730  \\
  C1997  &  J100058.01+023427.2  &   100  & 0 &  150.241741 &  2.574230  \\
  C2010  &  J100059.05+023508.9  &   237  & 0 &  150.246075 &  2.585822  \\
  C2011  &  J100059.12+023056.8  &    65  & 0 &  150.246349 &  2.515786  \\
  C2017  &  J100059.57+023736.2  &   102  & 0 &  150.248220 &  2.626733  \\
  C2023  &  J100059.80+023304.1  &    94  & 0 &  150.249200 &  2.551144  \\
  C2035  &  J100100.59+023305.2  &    74  & 0 &  150.252479 &  2.551469  \\
  C2048  &  J100101.61+023518.6  &    66  & 0 &  150.256737 &  2.588505  \\
  C2049  &  J100101.81+024111.7  &   169  & 0 &  150.257545 &  2.686591  \\
  C2052  &  J100102.19+023141.3  &   188  & 0 &  150.259141 &  2.528161  \\
  C2053  &  J100102.27+023034.5  &    82  & 0 &  150.259487 &  2.509599  \\
  C2059  &  J100102.59+023141.4  &   101  & 0 &  150.260808 &  2.528175  \\
  C2064  &  J100103.35+023229.7  &   225  & 0 &  150.263970 &  2.541583  \\
  C2067  &  J100103.63+024005.3  &    95  & 0 &  150.265129 &  2.668138  \\
  C2070  &  J100103.74+023053.2  &   254  & 0 &  150.265587 &  2.514788  \\
  C2071  &  J100103.77+023806.4  &    73  & 0 &  150.265729 &  2.635136  \\
  C2073  &  J100103.78+024212.3  &    57  & 0 &  150.265770 &  2.703438  \\
  C2078  &  J100104.26+023307.7  &    79  & 0 &  150.267783 &  2.552147  \\
  C2079  &  J100104.28+022806.8  &    59  & 0 &  150.267870 &  2.468572  \\
  C2086  &  J100104.58+023638.1  &   100  & 0 &  150.269091 &  2.610605  \\
  C2091  &  J100104.84+022859.6  &    69  & 0 &  150.270187 &  2.483222  \\
  C2092  &  J100104.95+023827.9  &    79  & 0 &  150.270650 &  2.641091  \\
  C2095  &  J100105.13+023004.3  &    70  & 0 &  150.271391 &  2.501219  \\
  C2096  &  J100105.55+023309.8  &   158  & 0 &  150.273162 &  2.552741  \\
  C2098  &  J100105.66+023238.1  &    82  & 0 &  150.273583 &  2.543919  \\
  C2102  &  J100106.07+023121.3  &    58  & 0 &  150.275329 &  2.522583  \\
  C2103  &  J100106.18+022707.5  &   216  & 0 &  150.275779 &  2.452083  \\
  C2118  &  J100107.13+023748.4  &   180  & 0 &  150.279733 &  2.630127  \\
  C2124  &  J100107.72+022736.9  &   212  & 0 &  150.282187 &  2.460250  \\
  C2129  &  J100108.43+023244.5  &    58  & 0 &  150.285141 &  2.545697  \\
  C2131  &  J100108.67+023022.7  &   189  & 0 &  150.286145 &  2.506325  \\
  C2136  &  J100108.99+022815.9  &   669  & 0 &  150.287491 &  2.471105  \\
  C2159  &  J100110.23+023127.2  &   138  & 0 &  150.292637 &  2.524230  \\
  C2160  &  J100110.43+023657.5  &    52  & 0 &  150.293475 &  2.615988  \\
  C2162  &  J100110.49+023226.4  &   218  & 0 &  150.293720 &  2.540666  \\
  C2174  &  J100111.30+022917.4  &   105  & 0 &  150.297116 &  2.488172  \\
  C2175  &  J100111.56+022840.8  &    59  & 0 &  150.298179 &  2.477999  \\
  C2177  &  J100111.61+023413.3  &    97  & 0 &  150.298383 &  2.570361  \\
  C2182  &  J100111.99+023442.3  &    78  & 0 &  150.299991 &  2.578422  \\
  C2184  &  J100112.06+024106.7  &  2464  & 0 &  150.300283 &  2.685202  \\
  C2193  &  J100113.52+022636.6  &    74  & 0 &  150.306341 &  2.443513  \\
  C2196  &  J100113.83+022645.2  &    61  & 0 &  150.307654 &  2.445905  \\
  C2198  &  J100114.08+022625.9  &    73  & 0 &  150.308675 &  2.440538  \\
  C2209  &  J100114.71+023518.3  &   122  & 0 &  150.311304 &  2.588441  \\
  C2224  &  J100115.21+024258.3  &    96  & 0 &  150.313412 &  2.716194  \\
  C2225  &  J100115.49+022858.4  &   422  & 0 &  150.314541 &  2.482913  \\
  C2229  &  J100116.13+022705.9  &    83  & 0 &  150.317229 &  2.451638  \\
  C2231  &  J100116.17+023241.8  &   133  & 0 &  150.317379 &  2.544952  \\
  C2235  &  J100116.59+022727.4  &   214  & 0 &  150.319124 &  2.457633  \\
  C2236  &  J100116.78+022830.5  &    57  & 0 &  150.319933 &  2.475161  \\
  C2247  &  J100117.23+023704.5  &    57  & 0 &  150.321795 &  2.617924  \\
  C2248  &  J100117.28+023308.6  &   137  & 0 &  150.322025 &  2.552413  \\
  C2249  &  J100117.36+023015.9  &    75  & 0 &  150.322333 &  2.504422  \\
  C2251  &  J100117.58+022657.5  &   216  & 0 &  150.323275 &  2.449305  \\
  C2254  &  J100117.70+024123.3  &    84  & 0 &  150.323750 &  2.689811  \\
  C2255  &  J100117.85+022654.2  &   297  & 0 &  150.324387 &  2.448411  \\
  C2257  &  J100117.95+022902.5  &   416  & 0 &  150.324829 &  2.484052  \\
  C2259  &  J100118.04+023610.3  &    77  & 0 &  150.325170 &  2.602877  \\
  C2265  &  J100118.57+022739.1  &   210  & 0 &  150.327395 &  2.460877  \\
  C2274  &  J100119.46+024307.4  &    56  & 0 &  150.331116 &  2.718727  \\
  C2283  &  J100119.92+023856.2  &   326  & 0 &  150.333000 &  2.648958  \\
  C2284  &  J100120.06+023443.7  & 10590  & 0 &  150.333600 &  2.578830  \\
  C2285  &  J100120.18+023403.2  &    53  & 0 &  150.334083 &  2.567569  \\
  C2289  &  J100120.41+022743.8  &   105  & 0 &  150.335066 &  2.462186  \\
  C2290  &  J100120.45+023834.2  &   183  & 0 &  150.335220 &  2.642844  \\
  C2298  &  J100120.85+022623.3  &    79  & 0 &  150.336912 &  2.439822  \\
  C2299  &  J100120.89+024001.7  &   107  & 0 &  150.337066 &  2.667155  \\
  C2302  &  J100121.29+023535.8  &   228  & 0 &  150.338733 &  2.593286  \\
  C2304  &  J100121.33+022648.6  &    89  & 0 &  150.338887 &  2.446855  \\
  C2308  &  J100121.82+023129.3  &    52  & 0 &  150.340920 &  2.524822  \\
  C2309  &  J100121.93+022814.7  &    70  & 0 &  150.341408 &  2.470775  \\
  C2313  &  J100122.02+023724.3  &   224  & 0 &  150.341775 &  2.623438  \\
  C2315  &  J100122.07+023405.5  &   235  & 0 &  150.341975 &  2.568211  \\
  C2330  &  J100123.17+023931.2  &    79  & 0 &  150.346549 &  2.658686  \\
  C2336  &  J100123.52+022618.2  &   120  & 0 &  150.348020 &  2.438402  \\
  C2338  &  J100123.76+023934.2  &   182  & 0 &  150.349020 &  2.659500  \\
  C2361  &  J100125.31+023527.5  &   131  & 0 &  150.355495 &  2.590986  \\
  C2362  &  J100125.36+023851.7  &   134  & 0 &  150.355683 &  2.647697  \\
  C2363  &  J100125.41+023145.1  &   472  & 0 &  150.355908 &  2.529205  \\
  C2370  &  J100125.95+023819.1  &    84  & 0 &  150.358162 &  2.638641  \\
  C2372  &  J100126.06+024149.5  &    70  & 0 &  150.358604 &  2.697108  \\
  C2374  &  J100126.28+023934.1  &    57  & 0 &  150.359500 &  2.659477  \\
  C2383  &  J100127.97+024029.3  &  1502  & 0 &  150.366570 &  2.674822  \\
  C2389  &  J100128.67+023311.8  &    68  & 0 &  150.369487 &  2.553286  \\
  C2403  &  J100129.48+023353.7A &   135  & 1 &  150.372860 &  2.564920  \\
  C2404  &  J100129.48+023353.7B &    51  & 1 &  150.372670 &  2.565030  \\
  C2405  &  J100129.48+023353.7  &   200  & 1 &  150.373020 &  2.564780  \\
  C2406  &  J100129.58+022716.3  &    56  & 0 &  150.373283 &  2.454552  \\
  C2407  &  J100129.83+023239.0  &   159  & 0 &  150.374291 &  2.544183  \\
  C2410  &  J100130.20+022856.1  &    81  & 0 &  150.375850 &  2.482263  \\
  C2414  &  J100130.44+024256.6  &   110  & 0 &  150.376850 &  2.715730  \\
  C2432  &  J100131.14+022924.7A &  4035  & 1 &  150.379770 &  2.490200  \\
  C2433  &  J100131.14+022924.7B &   407  & 1 &  150.377170 &  2.492450  \\
  C2434  &  J100131.14+022924.7C &   160  & 1 &  150.378670 &  2.491500  \\
  C2435  &  J100131.14+022924.7D &    56  & 1 &  150.379080 &  2.490900  \\
  C2436  &  J100131.14+022924.7  &  5377  & 1 &  150.379750 &  2.490210  \\
  C2437  &  J100131.14+022924.7E &    64  & 1 &  150.380540 &  2.489660  \\
  C2438  &  J100131.14+022924.7F &   189  & 1 &  150.380960 &  2.489240  \\
  C2439  &  J100131.14+022924.7G &   148  & 1 &  150.381580 &  2.488010  \\
  C2440  &  J100131.14+022924.7H &   431  & 1 &  150.382830 &  2.486830  \\
  C2443  &  J100131.36+022639.2  & 16120  & 1 &  150.380690 &  2.444240  \\
  C2444  &  J100131.36+022639.2A &  7485  & 1 &  150.380080 &  2.443780  \\
  C2445  &  J100131.36+022639.2B &  8746  & 1 &  150.381290 &  2.444690  \\
  C2446  &  J100131.36+022639.2C &   113  & 1 &  150.379120 &  2.444660  \\
  C2456  &  J100131.99+022807.8  &    59  & 0 &  150.383295 &  2.468847  \\
  C2460  &  J100132.14+023517.8  &    67  & 0 &  150.383916 &  2.588286  \\
  C2464  &  J100132.65+023232.2  &    70  & 0 &  150.386062 &  2.542280  \\
  C2470  &  J100133.58+022749.6  &   127  & 0 &  150.389945 &  2.463783  \\
  C2491  &  J100135.22+023109.0  &   279  & 0 &  150.396779 &  2.519177  \\
  C2493  &  J100135.25+023102.3  &   126  & 0 &  150.396912 &  2.517327  \\
  C2503  &  J100136.35+022751.3  &    67  & 0 &  150.401487 &  2.464272  \\
  C2505  &  J100136.46+022641.8  &   606  & 0 &  150.400880 &  2.445860  \\
  C2510  &  J100136.64+023639.2  &    81  & 0 &  150.402691 &  2.610897  \\
  C2512  &  J100136.93+023834.3  &   264  & 0 &  150.403908 &  2.642883  \\
  C2516  &  J100137.79+023054.9  &    78  & 0 &  150.407475 &  2.515272  \\
  C2517  &  J100137.85+023710.4  &    80  & 0 &  150.407745 &  2.619569  \\
  C2529  &  J100138.54+023736.8  &    66  & 0 &  150.410616 &  2.626888  \\
  C2535  &  J100139.36+023432.0  &   154  & 0 &  150.414000 &  2.575555  \\
  C2536  &  J100139.47+023351.5  &    82  & 0 &  150.414458 &  2.564311  \\
  C2541  &  J100139.85+023329.3  &   166  & 0 &  150.416041 &  2.558138  \\
  C2551  &  J100140.23+022735.8  &   166  & 0 &  150.417658 &  2.459952  \\
  C2552  &  J100140.28+023331.0  &    96  & 0 &  150.417845 &  2.558630  \\
  C2561  &  J100141.18+023250.8  &   105  & 0 &  150.421595 &  2.547463  \\
  C2566  &  J100141.42+023523.9  &    84  & 0 &  150.422616 &  2.589974  \\
  C2567  &  J100141.44+023157.2  &   652  & 0 &  150.422687 &  2.532566  \\
  C2569  &  J100141.50+023459.5  &   138  & 0 &  150.422954 &  2.583213  \\
  C2572  &  J100142.00+023049.6A &   501  & 1 &  150.425030 &  2.513800  \\
  C2573  &  J100142.00+023049.6B &   667  & 1 &  150.425040 &  2.513850  \\
  C2574  &  J100142.00+023049.6  &  1040  & 1 &  150.425170 &  2.513890  \\
  C2578  &  J100142.20+023503.3  &   132  & 0 &  150.425841 &  2.584261  \\
  C2586  &  J100142.49+023923.2  &   126  & 0 &  150.427070 &  2.656450  \\
  C2591  &  J100142.71+023331.1  &   268  & 0 &  150.427979 &  2.558658  \\
  C2602  &  J100143.86+023054.1  &    89  & 0 &  150.432787 &  2.515036  \\
  C2604  &  J100144.12+023724.5  &   160  & 0 &  150.433866 &  2.623483  \\
  C2610  &  J100144.55+023006.3  &    51  & 0 &  150.435625 &  2.501750  \\
  C2614  &  J100144.95+023835.5  &   136  & 0 &  150.437304 &  2.643194  \\
  C2616  &  J100145.15+023253.3  &    63  & 0 &  150.438124 &  2.548158  \\
  C2619  &  J100145.46+023236.8  &    54  & 0 &  150.439450 &  2.543577  \\
  C2620  &  J100145.55+023707.0  &    65  & 0 &  150.439829 &  2.618630  \\
  C2625  &  J100145.96+022854.0  &   154  & 0 &  150.441516 &  2.481683  \\
  C2626  &  J100145.96+023325.2  &   250  & 0 &  150.441537 &  2.557019  \\
  C2627  &  J100145.97+023948.0  &   262  & 0 &  150.441558 &  2.663333  \\
  C2628  &  J100146.22+022906.2  &   157  & 0 &  150.442604 &  2.485061  \\
  C2631  &  J100146.69+023251.6  &   122  & 0 &  150.444562 &  2.547688  \\
  C2633  &  J100146.79+023105.9  &   127  & 0 &  150.444979 &  2.518325  \\
  C2638  &  J100147.11+023916.6  &    68  & 0 &  150.446300 &  2.654625  \\
  C2641  &  J100147.18+023057.9  &    75  & 0 &  150.446600 &  2.516094  \\
  C2652  &  J100147.78+022913.6  &    96  & 0 &  150.449100 &  2.487111  \\
  C2662  &  J100149.61+023334.8  &  2202  & 0 &  150.456725 &  2.559691  \\
  C2673  &  J100150.67+023854.7  &    61  & 0 &  150.461129 &  2.648538  \\
  C2676  &  J100150.98+023541.2  &    61  & 0 &  150.462441 &  2.594791  \\
  C2678  &  J100151.15+023052.9  &    72  & 0 &  150.463129 &  2.514708  \\
  C2717  &  J100154.43+023240.0  &    98  & 0 &  150.476800 &  2.544444  \\
  C2718  &  J100154.47+023255.9  &   101  & 0 &  150.476983 &  2.548883  \\
\end{longtable}
%\begin{tablenotes}
%%\small
%\item {\textbf{Notes.}} {\it {Column 1}}: Source name used in the present project; {\it {Col. 2}}: Source name from \citet{schinnerer2010}; {\it {Col. 3}}: Integrated VLA flux density of the source (1.4 GHz) from \citet{schinnerer2010}; {\it {Col. 4}}: Classification between single- and multi-component source from \citet{schinnerer2010}, 0: single-component source, 1: multi-component source; {\it {Cols. 5, 6}}: Right ascension and declination (J2000) of the source from \citet{schinnerer2010}.
%\end{tablenotes}

\end{longtab}

\end{appendix}

\end{document}